\documentclass[superscriptaddress,
longbibliography,
amsmath,amssymb,preprint,
prf]{revtex4-1}

\usepackage[utf8]{inputenc}
\usepackage[T1]{fontenc}
\usepackage{nomencl}
\usepackage{graphicx}
\usepackage{dcolumn}
\usepackage{bm}
\usepackage{CJKutf8}
\usepackage{xcolor}
\usepackage{float}
\usepackage{amsmath}
\usepackage[pagewise]{lineno}
\linenumbers 

\begin{document}
\nolinenumbers
\title{Investigation on the Performance of a Torque-driven Undulatory Swimmer with Distributed Flexibility}

\author{Wenkang Wang (\begin{CJK*}{UTF8}{gbsn}王文康\end{CJK*})}

\affiliation{International Research Institute for Multidisciplinary Science, Beihang University, 100191 Beijing, China}
\affiliation{Max Planck Institute for Intelligent Systems, Heisenbergstraße 3, 70569 Stuttgart, Germany}

\author{Xu Chu (\begin{CJK*}{UTF8}{gbsn}初旭\end{CJK*})}
\email{xu.chu@simtech.uni-stuttgart.de}
\affiliation{Cluster of Excellence SimTech (SimTech), University of Stuttgart,
Pfaffenwaldring 5a, 70569 Stuttgart, Germany}

\date{\today} 

\begin{abstract}
The current study presents a systematic investigation of the locomotion performance of a swimmer with a wide range of parameter settings. Two-dimensional simulations with the immersed boundary method in the framework of Navier-Stokes equations are employed for the fluid-structure interaction analysis. Unlike most previous studies where the kinematics of the swimmer is predetermined, the locomotion of the current swimmer is the response of a single periodic torque applied on the anterior part. The effect of the distribution of body stiffness on swimming performance and propulsion generation is discussed with different pitch frequencies and amplitudes. An analysis of the phase-averaged vorticity field and thrust sequence is given to clarify the change of performance due to the variation of flexibility. This study demonstrates that body stiffness is a key factor that influences the performance of undulatory swimming when the pitch angle is low or moderate. The simple torque input of the current simulations provides a more direct and engineering-related insight for the future design of microrobotic swimmers.

\end{abstract}


\maketitle

\section{Introduction}

Undulatory swimming is a prevalent form of locomotion observed in a diverse range of biological organisms, extending from micrometer-scale spermatozoa to meter-scale cetaceans \citep{gazzola2014scaling, wu2022effects,quinn2021b,gao2023research}. Compared to traditional propulsion mechanisms such as rotational propellers, undulatory motion provides advantages including increased safety, diminished acoustic emissions, and adaptability to complex environmental settings replete with plant life and geological structures. These characteristics present significant opportunities for the implementation of undulatory swimming in robotic systems designed for environmental observation and remediation, as well as contributing to biomechanical investigations. 

Despite the creation of numerous robots employing undulatory motion \cite[]{esposito2012robotic,tangorra2010effect,wen2018understanding,wang2021effect}, these systems frequently fall short of replicating the swimming efficiency of their biological counterparts, primarily due to several prevailing challenges. A critical limitation restricting their efficacy is the constrained ability to tune stiffness. Unlike artificial systems, natural undulatory swimmers are able to actively and adaptively modify their body's stiffness distribution, thereby optimizing swimming performance across a variety of fluidic conditions and achieving superior propulsive velocity and energy conservation. Unfortunately, the examination of how stiffness distribution affects undulatory robots across various swimming modes remains an ongoing challenge.

Early efforts of computational fluid dynamics (CFD) simulations \cite[]{wu1971hydromechanics,zhu2007numerical,shoele2012leading} and experimental studies \cite[]{heathcote2007flexible,alben2012dynamics,dewey2013scaling} on flapping hydrofoils confirmed that adding flexibility can improve thrust or efficiency. \citet{hua2013locomotion} studied a free flapping plate with the immersed boundary (IB) method with the leading edge of the flexible plate forced to heave sinusoidally. They identified three distinct states of plate motion, that is, forward, backward, and irregular, which are determined by the heaving amplitude and the bending rigidity of the plate. It was shown that a suitable degree of flexibility can improve propulsive performance in the forward motion regime. \citet{quinn2014scaling} conducted experiments on flexible panels with heaving motions on the leading edge. They identified local maxima in the propulsive efficiency near resonant frequencies where the trailing edge amplitude is maximized. It was shown that the propulsive economy increases with higher flexibilities and slower swimming speeds at certain
Strouhal number conditions.

More recently, \citet{tytell2016role} studied the underlying physical mechanisms responsible for the influence of body stiffness using an immersed
boundary framework, with particular attention to the examination of wake structures. They found that the pressure distribution near the tail tip and the timing of vortex formation are in good agreement. It was also demonstrated that actuation at the resonant frequency dramatically increases propulsion efficiency. \citet{wang2020optimal} examined the effect of the non-uniform chordwise stiffness distribution on the self-propulsive performance of three-dimensional flexible plates.
It was shown that there is a common optimal stiffness for all plates to achieve the highest
cruising speed and efficiency. An analysis of force reveals that a greater deformation in the front part of the increasing rigidity pattern results in a greater thrust.

In addition to fundamental research on hydrofoils, introducing flexibility to the body, joints, or fins of fish-inspired robots also becomes an attractive concept in the field of robotics.
\citet{wang2021effect} designed a soft milli-swimmer inspired by the morphological properties of a larval zebrafish. Uniform stiffness and high swimming frequency (ranging from 60 to 100 Hz) were found to be beneficial for improving swimming speed and energy efficiency.
\citet{zhong2021a} created a dynamic model to explore the effects of adjusting stiffness on swimming performance. Their research showed that for fish-like robots to be as efficient as possible, the tension of the muscles must increase proportionally to the square of the swimming speed. This provides a simple way to adjust stiffness for different swimming tasks.
\citet{kwak2023development} developed a stiffness-adjustable articulated paddle and its application to a swimming robot. The thrust modulation caused by stiffness change is comprehensively studied by varying frequency and range of motion.

Despite the numerous studies on the impact of body stiffness on swimming performance, there is still a gap between theories and applications. One of the noticeable reasons is that most CFD simulations and experiments use predetermined kinematics for the whole or a portion of the swimmer. This is a straightforward way of conducting theoretical investigations. However, the kinematics of real-life fish or robots are the result of the interaction between internal and external forces and a complex fluid-structure system and are hard to prescribe. For example, a hydrofoil with pure heaving motion might be the most basic model in simulation and laboratory, but it is a nontrial task to build a real-life free-swimming robot with zero pitching angle since the pitching and heaving motion are strongly coupled in this fluid-structure system.  

To make contributions to the mentioned gap, we conduct numerical simulations with the immersed boundary method to examine the locomotion performance of a two-dimensional foil, focusing on various stiffness distributions throughout the body. Instead of defining the swimmer with prescribed motions, we actuate the swimmer with a single periodic torque applied to the anterior part. This allows us to focus on the locomotions that are achievable through this simple force input. In fact, magnetic torque is a widely used actuation approach in the area of soft robotics applications \cite[]{wang2021effect,wang2023concurrent}. In this way, we can link the analysis result of present simulations to actual robots that can be created in reality. In section \ref{sec:numerical}, we will introduce the numerical method and the swimmer model formulated in the current study. Section \ref{sec:results} is devoted to the presentation of simulation results, where the role that flexibility plays in propulsion and energy efficiency will be discussed. A conclusive remark will be given in section \ref{sec:conclusion}. 


\section{Numerical method}\label{sec:numerical}

\subsection{Governing equations and flow solver with immersed boundary method}

The immersed boundary formulation \citep{peskin2002immersed} of the problem presents the momentum and velocity of the coupled fluid-structure system in the Eulerian form while describing the deformation and elastic response of the immersed structure in the Lagrangian form. The 2-D incompressible Navier–Stokes equations describing the conservation of momentum and mass of fluid are solved for the fluid velocity $\mathbf{u}(\mathbf{x},t)$ and the pressure $p(\mathbf{x},t)$:

\begin{equation}
\nabla\cdot \mathbf{u}(\mathbf{x}, t)=0
\end{equation}

\begin{equation}
    \rho\left(\frac{\partial \mathbf{u}(\mathbf{x}, t)}{\partial t}+\mathbf{u}(\mathbf{x}, t) \cdot \nabla \mathbf{u}(\mathbf{x}, t)\right)=-\nabla p(\mathbf{x}, t)+\mu \Delta \mathbf{u}(\mathbf{x}, t)+\mathbf{f}(\mathbf{x}, t)
\end{equation}
where $\mathbf{f}(\mathbf{x}, t)$ is the force per unit area applied to the fluid by using the IBM and $\rho$, $\mu$ are the fluid density and dynamic viscosity respectively. Within the Lagrangian framework, $\mathbf{f}(\mathbf{x}, t)$ is calculated with

\begin{equation}
\mathbf{f}(\mathbf{X}, t)=\int \mathbf{F}(s, t) \delta(\mathbf{X}-\mathbf{X}(s, t)) \mathrm{d} r
\end{equation}

\begin{equation}
\mathbf{U}(\mathbf{X}(s, t), t)=\frac{\partial \mathbf{X}(s, t)}{\partial t}=\int \mathbf{u}(\mathbf{x}, t) \delta(\mathbf{X}-\mathbf{X}(s, t)) \mathrm{d} \mathbf{x}
\label{eq:U4}
\end{equation}

where $\mathbf{X}(s,t)$ is the Cartesian coordinate at time $t$
of the material point labeled by the Lagrangian parameter
$s$, and $\mathbf{F}(s,t)$ is the force per unit area imposed onto the fluid by elastic deformations in the immersed structure as
a function of the Lagrangian position $r$, and time $t$.
The flow motions and structure deformation are updated iteratively.
The fluid-structure interaction is simulated using the open-source code {\it{IB2d}} \cite{battista2017ib2d,battista2018ib2d}, which has been testified and validated by various authors \citep{wang2023concurrent}. {\it{IB2d}} uses finite difference approximations to discretize the Navier–Stokes equations on a fixed grid with the Fast Fourier transform (FFT) algorithm. FFT algorithm shows a significant advantage in the computational efficiency over the iterative methods and therefore has been widely deployed in single- \citep{chu2019direct} and multi-phase solvers \citep{chu2020turbulence,liu2023large}.  

\subsection{The modeling of the undulatory swimmer}

In the current simulations, the swimmer is modeled as a mass-less fiber. 
In the {\it{IB2d}} framework, we employed virtual springs and virtual beams to connect successive points along the angular fish.
The deformation of the fiber is computed with the combination of spring and beam models. The damped spring model reads as:

\begin{equation}
\mathbf{F}_{spring}=k_s(|\Delta \mathbf{X}|-R_L)\Delta \mathbf{X}/|\Delta \mathbf{X}|
\end{equation}
where $k_s$ is the spring stiffness, $R_L$ is the spring's resting length.
And $\Delta\mathbf{X}$ is the distance between two successive Lagrangian points.

In the noninvariant beam model, the bending deformation force is modeled as:

\begin{equation}
\mathbf{F}_{beam}=k_{b} \frac{\partial^{4}}{\partial s^{4}}\left(\mathbf{X}(s, t)-\mathbf{X}_{b}(s, t)\right)
\end{equation}

and $k_b$ is the beam stiffness. The resistance to bending is calculated between 3 successive Lagrangian points. $\mathbf{X}(s, t)$ is the current Lagrangian configuration at time $t$ where $\mathbf{X}_{b}(s, t)$ is the preferred configuration at time $t$.

\begin{figure}
\includegraphics[scale=0.6]{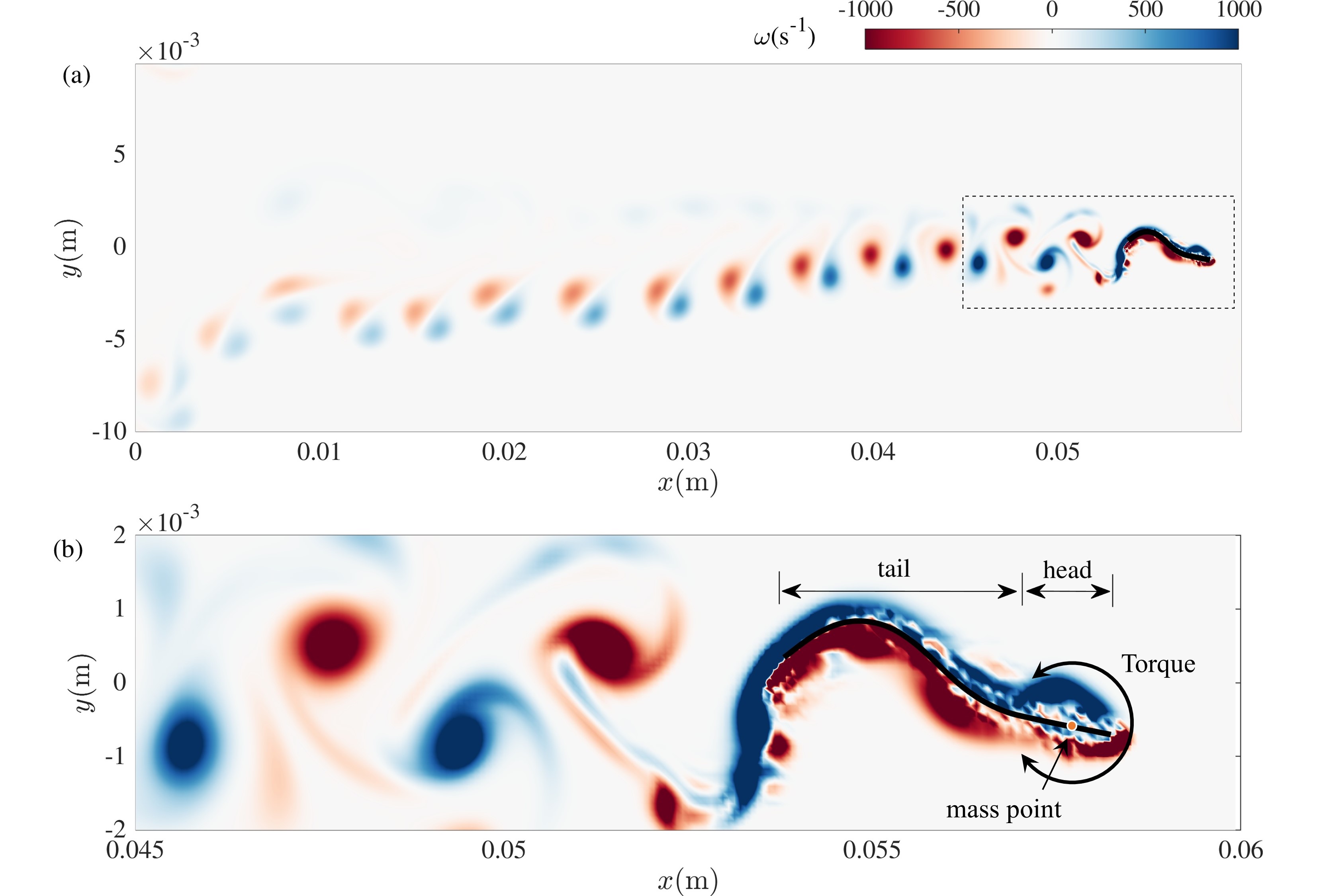}
\caption{(a) An illustration of the computational domain. The color contour shows the vorticity in the wake of the swimmer (pitch angle $\theta=35^\circ$, frequency $f=80$Hz, flexibility coefficient $a=3$). (b) A close-up view of the dashed-box region in (a). The design of the swimmer model is depicted.}
\label{fig:fish_model}
\end{figure}

The computational domain and the swimmer configuration are demonstrated in figure \ref{fig:fish_model}. Note that in our current simulations, we established a direct correspondence between our code units and the units in the real world with a conversion factor of 1 for each variable, including length, time, force, etc. This approach allows us to seamlessly translate our simulation results into real-world values without scaling. For example, when representing lengths, each unit in our simulation directly equates to one meter in the physical world. This direct alignment simplifies the interpretation of our findings, enabling more straightforward and intuitive analysis and comparison with experimental data.

For all cases in the current study, the computational domain is $\Delta X\times\Delta Y=0.06$ m $\times$ 0.02 m (see figure \ref{fig:fish_model}a).
The swimmer has a chord length of $C=5$ mm, and consists of a rigid head and a flexible tail, as shown in figure \ref{fig:fish_model} (b). The length ratio between the rigid head and the flexible tail is 1:3.  For the head, the bending stiffness $k_h$ is set at an extremely high value ($1\times10^{9}$ $\mathrm{N\cdot m^2}$) so that the deformation of the head part is negligible. The stiffness of the flexible body follows an exponential decreasing trend, i.e.,
\begin{equation}
    k_t=k_{t}^0\mathrm{exp}(-a(s-s_1)/C)
    \label{eq:kb}
\end{equation}
where $s_1$ is the Lagrangian coordination for the joint between the head and body, coefficient $a$ is a variable that controls the decreasing rate, and the bending stiffness $k_{t}^0$ is set to $3\times10^{-11}\mathrm{N\cdot m^2}$, similar to the properties of the natural organism \cite[]{wang2021effect,van2015body}. In fact, the configuration of the swimmer is inspired by the larval zebrafish ($Danio$ $rerio$), which has a similar body length and head-tail ratio\cite[]{kimmel1995stages}. The locomotion of zerbrafish larvae has been extensively investigated in experiments \cite[]{mueller2004,van2015body}. This provides us with data for validation of the simulation results.

To generate undulating motions, we apply distributive external force on the fiber on the anterior 25\% length of the swimmer, i.e., the head part. The force is formulated as a target force that pulls Lagrangian point $X$ towards a target position $\mathbf{X}_T$, that is, 
\begin{equation}
\mathbf{F}_{T}=-k_T(\mathbf{X}-\mathbf{X}_{T})
    \label{eq:target_force1}
\end{equation}

\begin{equation}
    \mathbf{X}_{T}=\mathbf{T}(\mathbf{X}_c)\mathbf{R}[\theta\mathrm{sin}(2\pi ft)]\mathbf{T}(-\mathbf{X}_c)\mathbf{X}
    \label{eq:target_force2}
\end{equation}
where $k_T$ is the penalty coefficient that works as the stiffness of a virtual spring connecting $\mathbf{X}$ and $\mathbf{X}_T$.  The target position $\mathbf{X}_T$ is formulated by rotating the current position $\mathbf{X}$ around the center $\mathbf{X}_c=(X_c,Y_c)$, which is located at the middle of the head part, i.e., 0.125$L$ from the front tip. $\mathbf{T}$ and $\mathbf{R}$ in Eq.\ref{eq:target_force2} are 2D translation matrix and rotation matrix, respectively, which are defined as follows,

\begin{equation}
    \mathbf{T}(X_c,Y_c)=
    \begin{bmatrix}
1 & 0 & X_c\\
0 & 1 & Y_c\\
0 & 0 & 1
\end{bmatrix}
\end{equation}

\begin{equation}
    \mathbf{R}(\theta_t)=
    \begin{bmatrix}
\mathrm{cos}(\theta_t) & -\mathrm{sin}(\theta_t) \\
\mathrm{sin}(\theta_t) & \mathrm{cos}(\theta_t)
\end{bmatrix}
\label{eq:rotation_matrx}
\end{equation}

Note that an expansion to a third position of $\mathbf{X}$ is required to apply translation matrix, i.e., $\mathbf{X}=[X,Y]^{\mathrm{T}}\rightarrow  [X,Y,1]^{\mathrm{T}}$. The distributive target force defined by Eq.\ref{eq:target_force1}-\ref{eq:rotation_matrx} is antisymmetric about $\mathbf{X}_c$ and forms a torque that bends the head to a periodic pitch angle $\theta_t=\theta\mathrm{sin}(2\pi ft)$ (see figure \ref{fig:fish_model}b).

Unlike the majority of prior studies, where the pitching and heaving motions of the fish were specified by directly programming a part or the whole body's kinematics \cite[]{tytell2016role,floryan2020distributed,van2013optimal,smits2019undulatory,quinn2021b,zhong2021a}, the locomotion of the present swimmer is completely determined by the reaction of the body to the torque applied to the head. The only constraints are the target pitch angle $\theta$ and the actuation frequency $f$. The heaving motions of the leading edge, which is usually predetermined in other studies, are now an output of the fluid-structure system. The current actuation method is similar to the real-life scenario of an untethered swimmer where the kinematics of the body is only the response of body force to the fluid-structure interaction. Therefore, the current study has a closer connection to biomimetic and robotic applications, particularly those that use magnetic actuation \cite[]{wang2021effect}.

At this point, even though we are applying external forces and a high degree of stiffness to the head section, the head and tail still move in a symmetrical undulating motion, and no directional thrust is produced. This is partly attributed to the assumption of the immersed boundary method that the structure nodes have no mass. In nature, however, the mass distribution of a larval zebrafish is strongly asymmetrical along the body length with a mass center located at the anterior 25\%-29\% of the body \cite[]{van2015body}. Therefore, we introduce a virtual mass point to our swimmer model to break the head-tail symmetry. At the mass point, a time-dependent force is applied to mimic the effect inertia, which is proportional to the acceleration rate of the local Lagrangian node, that is,



\begin{equation}
\mathbf{F}_{mass}=-Ma
\label{eq:fmass}
\end{equation}
The mass point is situated in the middle of the head area, which is also the center of the external torque $\mathbf{X}_c$ (see figure \ref{fig:fish_model}). The mass value is set as the mass of a fluid particle of size $\Delta s$, the interval of Lagrangian points. Tests show that the adulatory motion of the fiber is slightly affected by mass value $M$. However, swimmers with a large mass point, $M$, are more difficult to accelerate and take longer to reach a steady swimming state. We have chosen a small value that is sufficient to break the head-tail symmetry and does not significantly affect the acceleration process.

\subsection{Convergence test}
In this study, we discretize the computation domain with a uniform grid spacing $\Delta x=\Delta y=\Delta X/768=\Delta Y/256\approx$0.0078mm. The grid spacing of the Lagrangian structure is set to $\Delta s=\Delta x/2$. The time step is $\Delta t=2\times10^{-6}$ s and the CFL condition is less than 0.01. A test of convergence is performed to ensure that the swimmer's motion is not influenced by the current mesh resolution.
 
\begin{figure}
\includegraphics[scale=0.16]{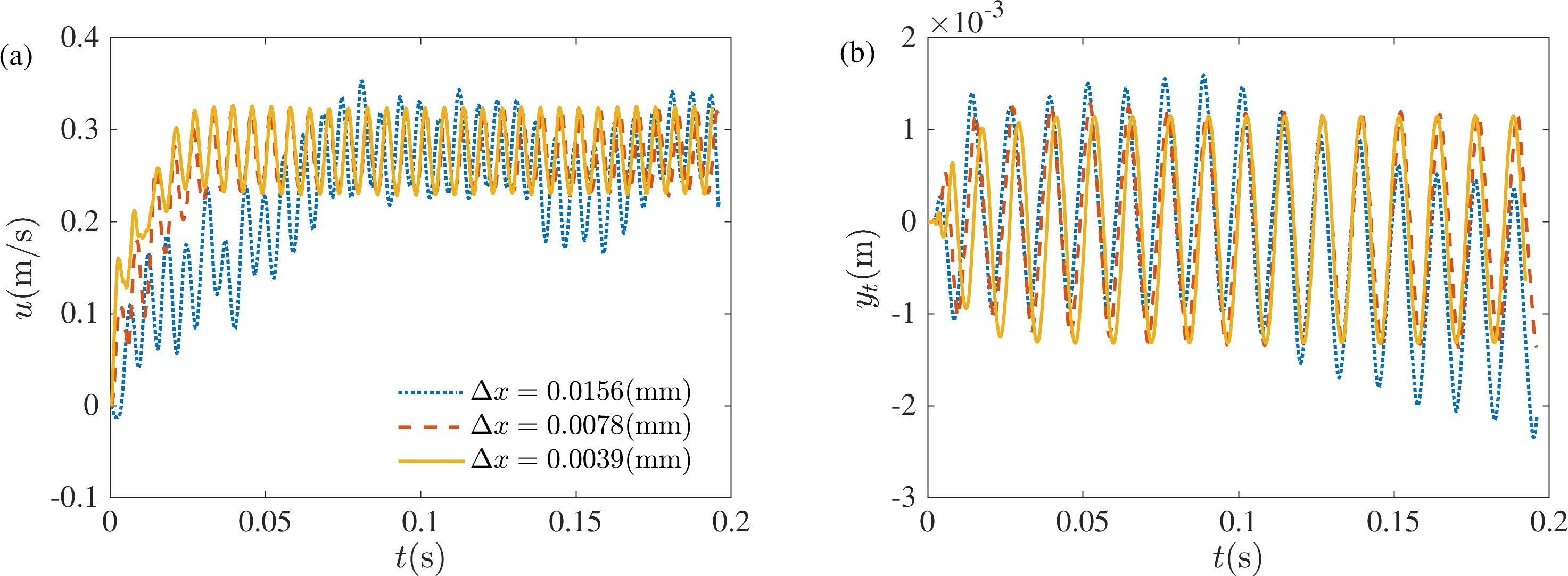}
\caption{Mesh convergence test with three different mesh sizes $\Delta x=0.0156$ mm, $0.0078$ mm and $0.0039$ mm. (a) The time evolution of swimming speed $u$; (b) The time evolution of the trailing edge amplitude $y_t$. The swimmer has a pitch angle $\theta=35^\circ$, actuation frequency $f=80$Hz, and flexibility coefficient $a=3$.  }
\label{fig:convergence}
\end{figure}

Figure \ref{fig:convergence} compares the swimming speed and trailing edge amplitude of a swimmer ($\theta=35^\circ$, $f=80$, flexibility coefficient $a=3$, see figure \ref{fig:fish_model}a for illustration) with three different resolutions. When the grid space is $\Delta x=0.0156$mm, the swimmer has a slower acceleration, and it is difficult to reach a stable swimming state. For the two finer mesh sizes, $\Delta x=0.0078$mm and $\Delta x=0.0039$mm, the time evolution of the swimming speed and the amplitude of the trailing edge are almost identical, indicating the convergence of the computation. Therefore, we performed all of the simulations below with intermediate resolutions.

\begin{table}
\centering

\begin{tabular}{|c|c|}
\hline
\textbf{Parameters} & \textbf{value} \\
\hline\hline
\multicolumn{2}{|c|}{Input parameters}\\
\hline
Oscillation frequency   &  $f\in[30,100]$Hz  	\\
Pitch angle	&   $\theta\in[15,55]^\circ$   	\\
Bending stiffness &  $k_t\in[0.3,3]\times10^{-11}\mathrm{N\cdot m^2}$\\

Chord length  	&     $C=5\times10^{-3}$m  	\\
Span  	&   $s=0.6\times10^{-3}$m    	\\
Density of fluid &     $\rho=10^3\mathrm{kg/m^3} $ \\
Visocity of fluid & $\nu=10^{-6}\mathrm{m^2/s}$\\
\hline
\multicolumn{2}{|c|}{Output parameters}\\
\hline
Swimming velocity &$\overline{u}\in[0.002,0.4]$m/s\\
Amplitude of trailing edge & $A\in[0.6,2.1]\times10^{-3}$m\\
Reynolds number & $Re=\overline{u}C/\nu\in[10,1850]$\\
Strouhal number & $St=fA/\overline{u}\in[0.25,4.52]$\\
Cost of transport & $\mathrm{CoT}=\overline{P}/(M\overline{u})\in[100,700]$J/(kg$\cdot$m)\\
\hline
\end{tabular}
\caption{A summary of simulation parameters}
\label{table1}
\end{table}

\section{Results}\label{sec:results}

\subsection{Parameter space generated by Sobol sequence}

\begin{figure}
\includegraphics[scale=0.15]{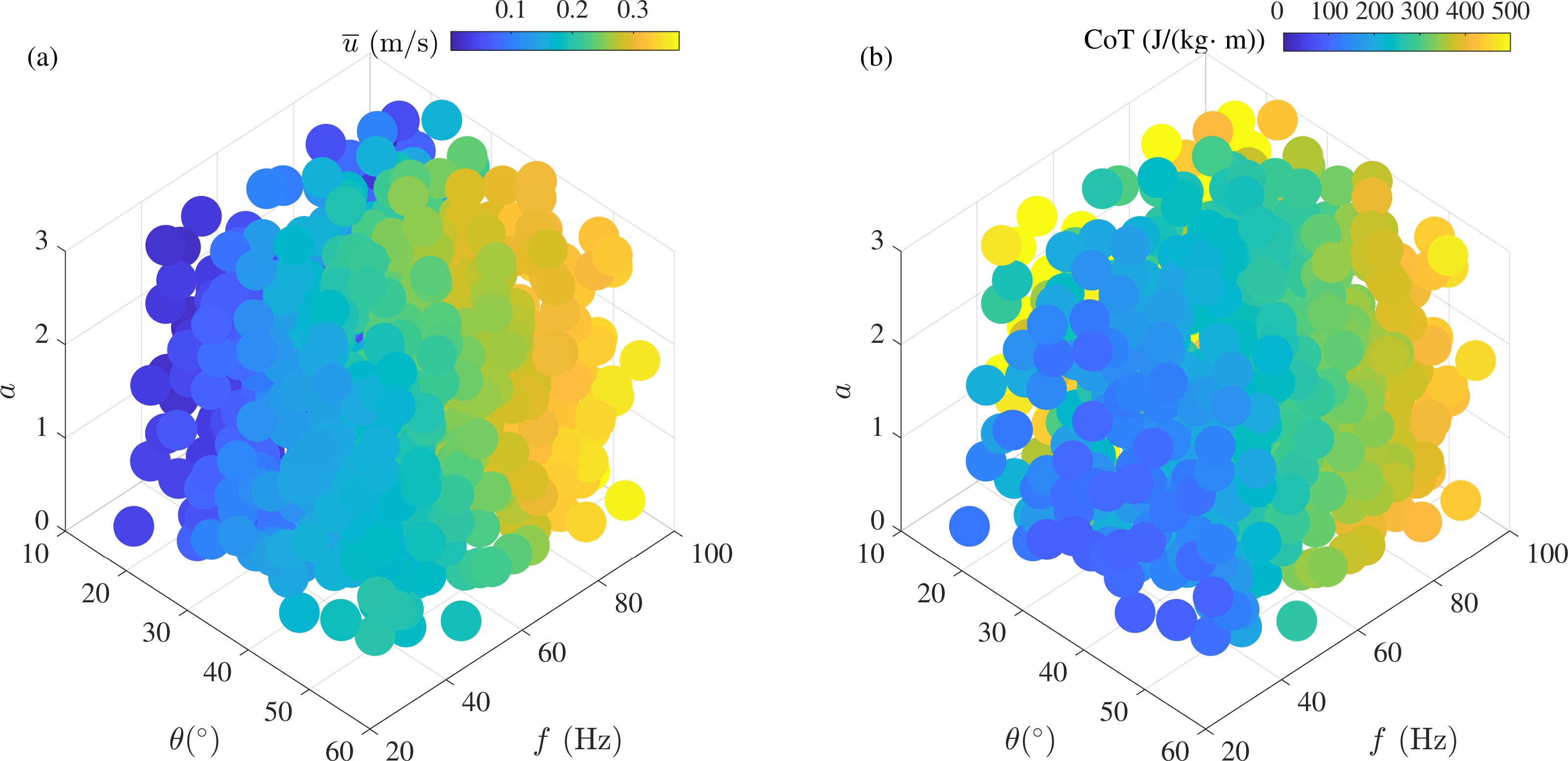}
\caption{3D parameter space being sampled by Sobol sequences, the color legend indicates mean streamwise velocity $\overline{u}$ and torque $T$.}
\label{fig:sobol}
\end{figure}

In the current study, the pitch frequency $f$, pitch angle $\theta$, and uniformity coefficient $a$ are the key control parameters of the swimming performance. To facilitate a more efficient exploration of the parameter space, we use the Sobol sequence to sample the parameter space $a\in[0, 3]$, $f\in[30, 100]$ Hz, $\theta\in[15,55]^\circ$. This approach not only facilitates a more efficient exploration of parameter spaces compared to uniform grid sampling but also enhances the convergence rates in numerical integration and optimization problems. The deterministic nature of Sobol sequences also allows reproducible results that are possible for validation. 

A total of 600 samples were generated that were taken as input parameters to the simulation. All the parameters used in current simulations are summarized in table \ref{table1}. It is observed that, although the input parameters are confined to a limited range, the output parameters vary significantly between cases. For instance, the swimming speed and Reynolds number span three orders of magnitude, while the Strouhal number fluctuates by more than one order. The current study establishes links between input parameters, such as the actuation frequency $f$ and the pitch angle $\theta$, and the speed of swimming $\overline{u}$ and the amplitude of the trailing edge $A$. This is especially relevant as certain factors, such as $Re$ and $St$, cannot be predetermined. 

An overview of all cases is provided in figure \ref{fig:sobol} where the mean swim speed $\overline{u}$ and the cost of transport (CoT) are illustrated in colors. Figure \ref{fig:sobol} gives a general idea about the distribution of samples in the parameter space and the swim performance of different combinations of input parameters. In the following sections, the variables that influence swimming performance will be explained in more detail.





\subsection{The effect of flexibility on swimming performance}

The performance of the fish is reflected by two variables, namely steady swimming velocity $\overline{u}$ and cost of transport (CoT). The steady swimming velocity $\overline{u}$ and CoT are calculated by averaging over 6 periods after the swimmer reachs a stable swimming. CoT is a common criterion for assessing locomotion-related energy efficiency \cite[]{voesenek2020experimental}, which is defined as 
\begin{equation}
    \mathrm{CoT}=\overline{P}/(M\overline{u})
    \label{eq:cot}
\end{equation}
where the overline represents time-averaging. $\overline{P}$ is the mean power input that is calculated by integrating the power of target forces on the swimmer, that is, 

\begin{equation}
    \overline{P}=\overline{\int_L F_{T}(s,t)u(s,t)\mathrm{d}s}
    \label{eq:P}
\end{equation}
where $L$ represents the domain of Largrangian points. In Eq.\ref{eq:cot}, $M$ is the mass of the swimmer. Since the swimmer is massless under the framework of immersed boundary method (except for the extremely small virtual mass at the head), to make it possible to compare to the CoT of real fish, we use the added fluid mass when computing the CoT, which is estimated to be $M_{\mathrm{added}}=4\rho_{2D}C\Delta s$ base on the Dirac-delta support function. Since the velocity of the local Lagrangian node $u$ is estimated by the velocity of the fluid points that support it (Eq.\ref{eq:U4}), the movement of the swimmer can be seen as the result of the transport of the added fluid due to the energy input. Therefore, it is reasonable to use the mass of the added fluid for the calculation of CoT.

 \begin{figure}
\includegraphics[scale=0.12]{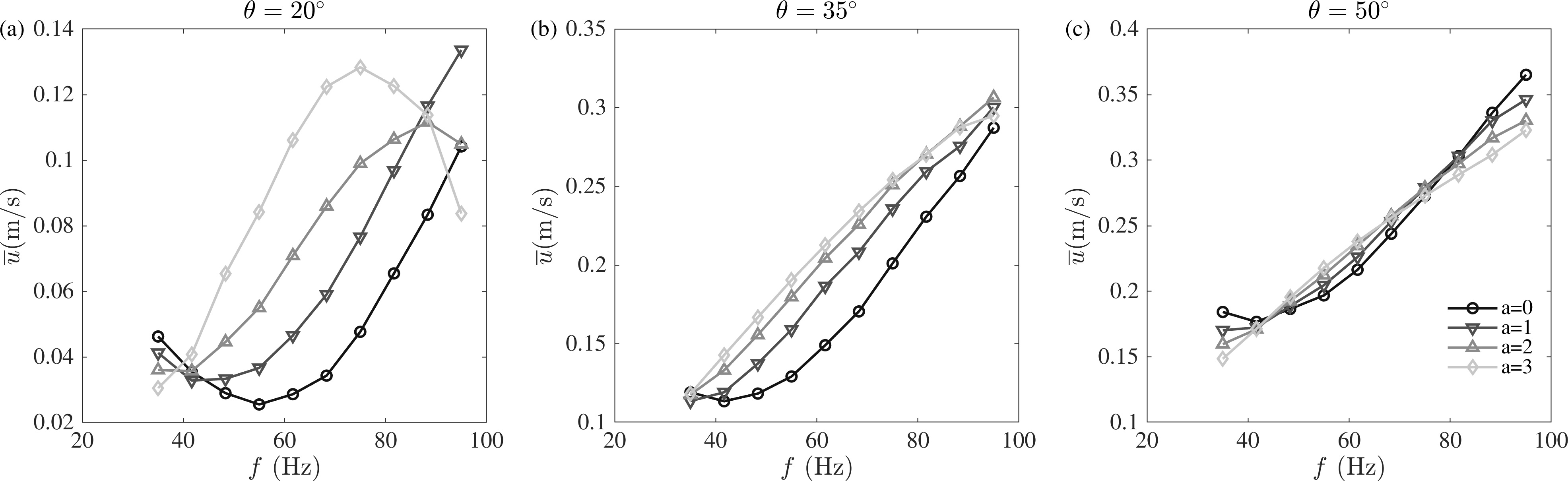}
\caption{Mean swimming speed $\overline u$ as a function of frequency for three pitch angles (a) $\theta=20^\circ$, (b) $\theta=35^\circ$ and (c) $\theta=50^\circ$.}
\label{fig:u_vs_a}
\end{figure}

 \begin{figure}
\includegraphics[scale=0.12]{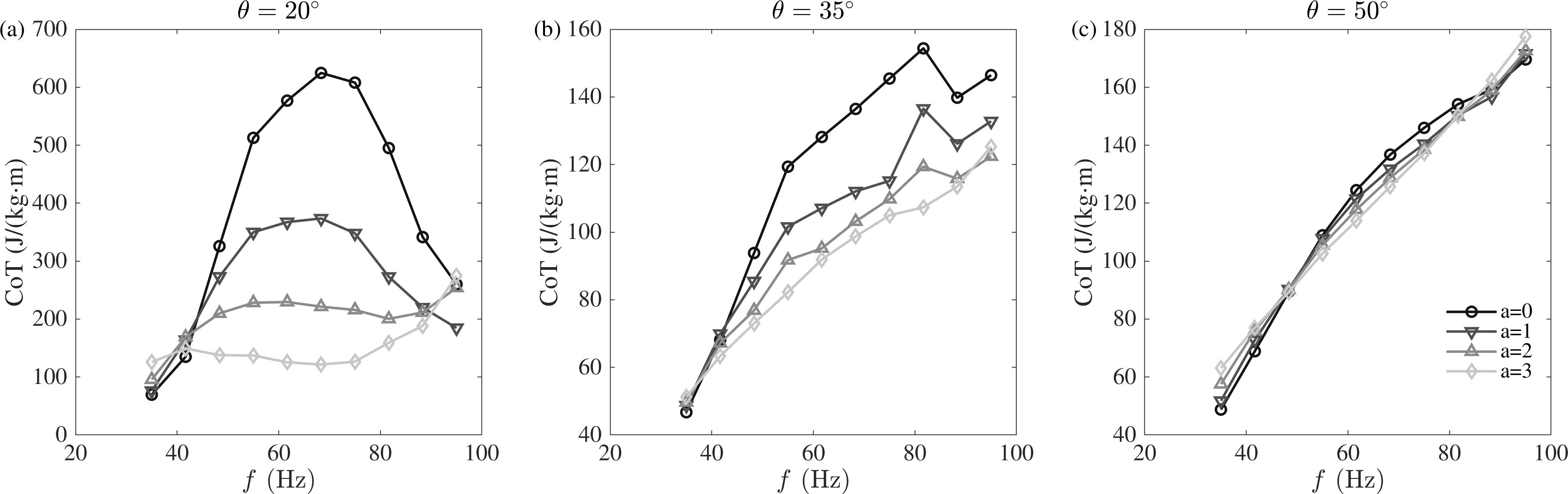}
\caption{CoT as a function of frequency for three pitching angles (a) $\theta=20^\circ$, (b) $\theta=35^\circ$ and (c) $\theta=50^\circ$.}
\label{fig:cot_vs_a}
\end{figure}

To show the role played by the pitching amplitude, the pitching frequency, and body stiffness more clearly, we show $\overline{u}$ and CoT as a function of $f$ at selected pitching angle $\theta$ and distribution coefficient $a$ in figure \ref{fig:u_vs_a} and figure \ref{fig:cot_vs_a}, respectively. Note that the points on the curves are linearly interpolated from the samples shown in figure \ref{fig:sobol}. 

For all three pitch angles in figure \ref{fig:u_vs_a}, a high pitching frequency is generally favorable to reach a high swimming speed $\overline{u}$. This is within expectations since a higher pitching frequency is associated with a larger pitching torque and input power. However, there are cases where raising $f$ leads to a decrease in swimming speed. For example, at a low pitch angle $\theta=20^\circ$, the steady swimming speed $\overline {u}$ of a soft tail swimmer ($a=3$) decreases with $f$ when $f>70$Hz. This implies that the input energy is wasted on extra deformations or lateral movements rather than on generating thrust. The existence of extreme points on the curves in figure \ref{fig:u_vs_a} suggests transitions of swimming dynamics, which will be further inspected in later sections.

It is noticeable that the variation of the flexibility distribution has a considerable effect on $u$ for small and intermediate pitch angles. For the cases of high pitch angle ($\theta=50^\circ$), there is only a slight variation in $\overline{u}$ between different stiffness distributions. For low and medium pitch angles ($\theta=20, 35^\circ$), however, larger values $a$ are more beneficial in terms of swimming speed. The reason might be that when the angle of attack is small, the flexibility is more relevant to the shape of the body wave, especially the trailing edge amplitude. In later sections, it will be shown that swimmers with a soft tail (i.e. larger $a$) generate a larger trailing edge amplitude, which is proportional to thrust generation \cite[]{floryan2020distributed}. 

Figure \ref{fig:u_vs_a}(a) shows that a soft tail swimmer ($a=2,3$) has the advantage in swimming speed in the frequency range of $f=40-80$Hz at a low pitch angle. Meanwhile, a stiff-tail swimmer ($a=0, 1$) may achieve the same swimming speed by oscillating faster. However, this is at the cost of more energy, which is demonstrated by the CoT curves in figure \ref{fig:cot_vs_a} (a). Soft tail swimmers ($a=2$ and 3) are significantly more efficient than those with stiffer tails in the frequency range $f=40-80$Hz, with optimal frequency identified at 68Hz and 81Hz, respectively. As the pitching angle grows ($\theta=35^\circ,50^\circ$), the discrepancy between the different stiffness distribution cases decreases, and the CoT becomes monotonically related to the frequency. In such cases, the energy cost to maintain the speed of swimming increases as $f$ increases. 

It is also observed that a medium pitch angle $\theta=35^\circ$ is more efficient than the cases of low $\theta=20^\circ$ and high $\theta=50^\circ$ angle in the same $f$. When the pitch angle is too low, the amplitude of the trailing edge is not maximized. Moreover, for the stiff-tail cases, the body wave is also not fully developed. However, when the pitch angle is too high, energy may also be wasted on extra drag and lateral movement. This is consistent with observations on real larval zebrafish and biomemetic robots \cite[]{mueller2004,wang2021effect}.

In this section, we identified that the stiffness distribution plays an important role in reducing energy costs when the pitching angle is low. In high frequency and large angle pitching, the stiffness distribution is less important. To further understand this phenomenon, the role that flexibility plays in propulsion generation will be explored in the following sections.  

\subsection{The effect of flexibility on propulsion generation}

\begin{figure}
\includegraphics[scale=0.12]{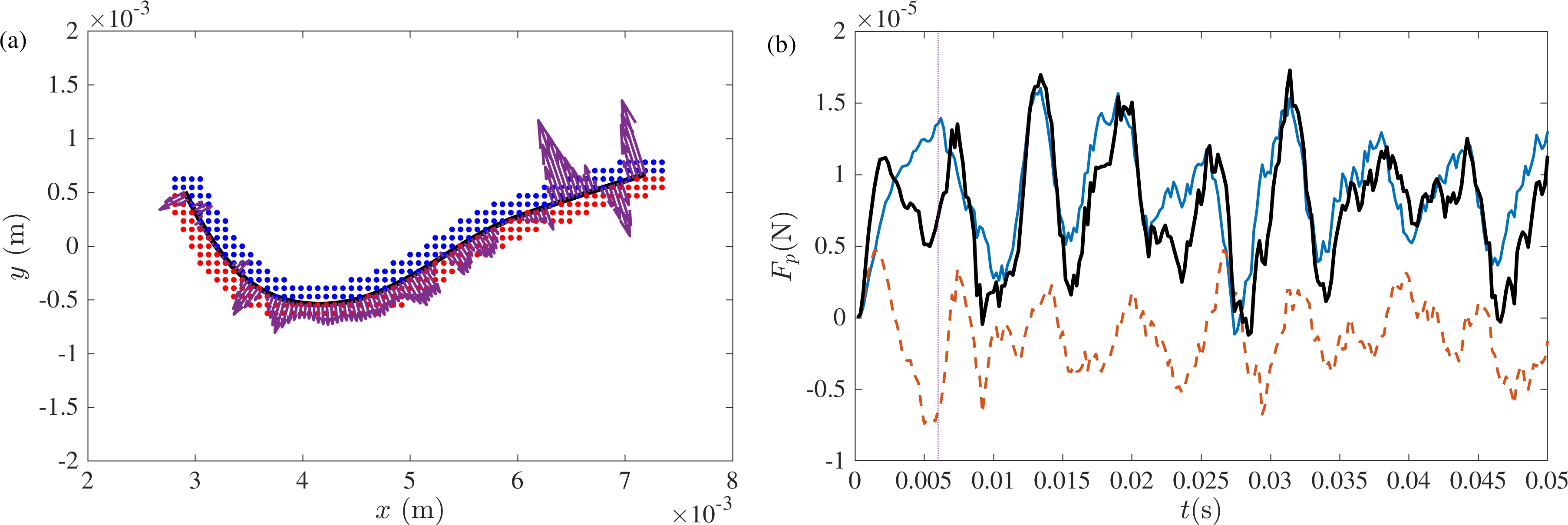}
\caption{(a) An illustration for the integration of pressure force based on its surrounding Eulerian points. The configuration of the swimmer is the same as in figure \ref{fig:fish_model}, i.e., pitch angle $\theta=35^\circ$, frequency $f=80$Hz, flexibility coefficient $a=3$. Blue and red points are the support Eulerian points on the upper and lower sides, respectively. The arrows indicate the direction and magnitude of local pressure force $\mathbf{f}_p$.  (b) Time series of streamwise thrust/drag from the tail (blue solid line) and head (red dashed-line). The black solid line represents the thrust of the whole body. The vertical dot line indicates the position of (a) on the time-series.}
\label{fig:thrust}
\end{figure}

To look further into the dynamics of swimming performance, we computed the thrust of the swimmer by integrating pressure along the body. For each Lagrangian point on the swimmer, the pressure force is estimated as a weighted integral of its surrounding Eulerian points:
\begin{equation}
    \mathbf{f}_p(\mathbf{X}(s, t), t)=\int p(\mathbf{x}, t) \delta(\mathbf{X}-\mathbf{X}(s, t)) \mathrm{sgn}(Y-Y(s,t))\cdot \mathbf{n} \mathrm{d} \mathbf{x}
    \label{eq:Fp}
\end{equation}
where $\mathbf{n}$ is the normal vector of the local segment on the swimmer. Figure \ref{fig:thrust} shows an illustration of the integration in Eq.\ref{eq:Fp}, where the Eulerian grids on each side of the swimmer are labeled by different colors. The arrows show the direction and magnitude of the net pressure force at local Lagrangian points. At the illustrated moment, the majority of the pressure forces $\mathbf{f}_p$ on the tail have a streamwise component that is positive, thus providing thrust for the swimmer. On the head, the pressure forces are mostly pointing in the opposite direction, indicating drag caused by pressure.

We assess the pressure force contribution from the tail and head by integrating the streamwise component of the pressure force $\mathbf{f}_p$ in both sections. Time sequences of the integrals $F_p$ are shown in figure \ref{fig:thrust}(b). The thrust of the swimmer is generated mainly from the tail part, as evidenced by the fact that the total force on the tail is positive, while the force on the head is mostly negative. This thrust is used to counteract friction drag and allow the swimmer to accelerate. 

\begin{figure}
\includegraphics[scale=0.12]{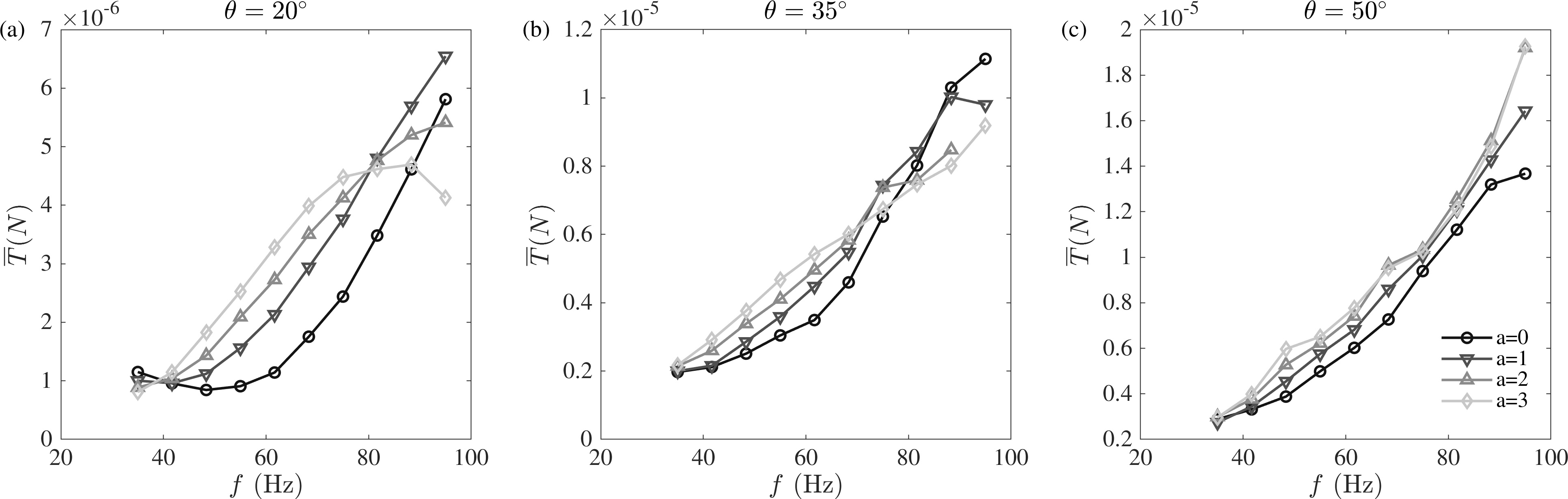}
\caption{Mean thrust $\overline T$ as a function of frequency for three pitch angles (a) $\theta=20^\circ$, (b) $\theta=35^\circ$ and (c) $\theta=50^\circ$.}
\label{fig:T_f}
\end{figure}

The net thrust of the swimmer $T$ can be defined as the integral of the streamwise pressure force along the body length. The change in net thrust $T$ with frequency $f$, pitch angle $\theta$ and flexibility coefficient $a$ is illustrated in figure \ref{fig:T_f}. For a low pitch angle ($\theta=20^\circ$), the soft tail configurations ($a$=2,3) generate more thrust than stiff tails ($a$=0,1) in the range frequency range of the range of $f=40-80$Hz. This range overlaps with the frequencies where the soft tail swimmers achieve a higher swim speed (see figure \ref{fig:u_vs_a}a). For higher pitch angles ($\theta=35^\circ$ and $50^\circ$), the impact of tail flexibility on thrust is less significant. The thrust increases monotonically with the actuation frequency.

\begin{figure}
\includegraphics[scale=0.12]{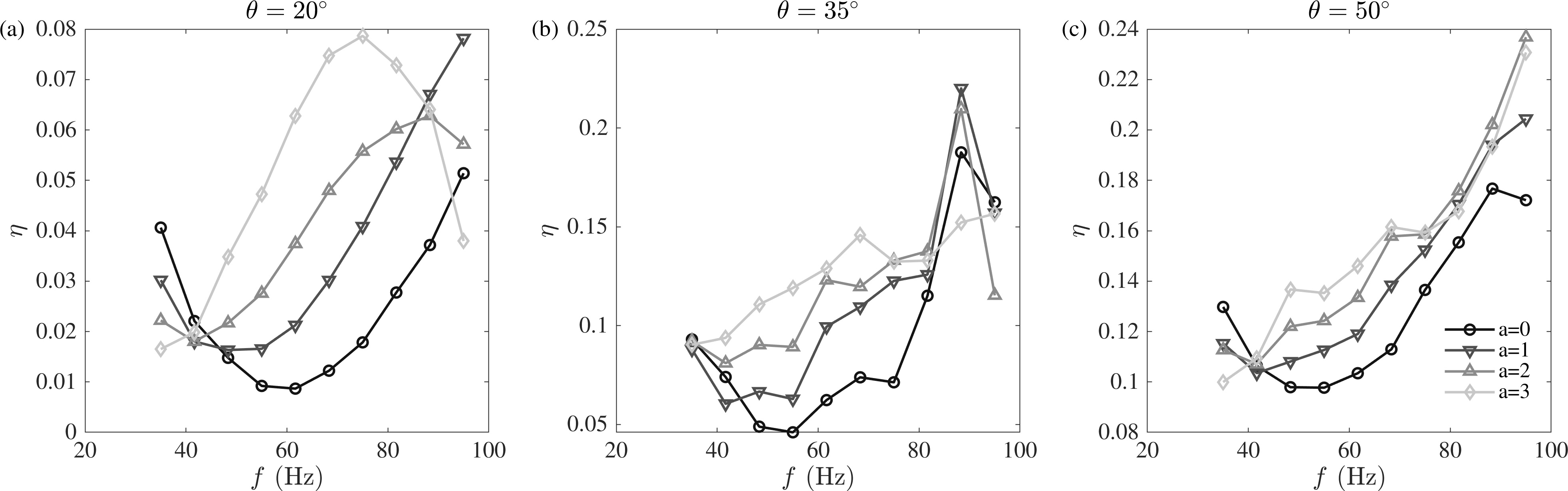}
\caption{Propulsive coefficient $\eta$ as a function of frequency with three pitch angles (a) $\theta=20^\circ$, (b) $\theta=35^\circ$ and (c) $\theta=50^\circ$.}
\label{fig:eta}
\end{figure}

In addition to the amplitude of thrust $T$, we computed the propulsive coefficient,

\begin{equation}
    \eta=\overline{T}\overline{u}/\overline{P}
\end{equation}
where $\overline{P}$ is the time-averaged input power as defined in Eq.\ref{eq:P}. The propulsive coefficients reflect the percentage of the input energy that is converted into thrust work. The variation of $\eta$ with $f$ is shown in figure \ref{fig:eta}. Generally, the soft tail configurations ($a=2,3$) have higher propulsive efficiency than the stiff tail ones, especially when the actuation frequency falls in the range of $40\mathrm{Hz}<f<80\mathrm{Hz}$. When the actuation frequency is higher than 80Hz, the difference of $\eta$ between flexibility distribution is not significant. This is also consistent with the result of CoT in figure \ref{fig:cot_vs_a}.

In this section, we present additional proof of the effect of flexibility on thrust generation, which partially explains the swimming capability of various configurations. In the following section, the effects of flexibility on the kinematics will be thoroughly examined to gain further understanding.

 \begin{figure}[ht!]
\includegraphics[scale=0.12]{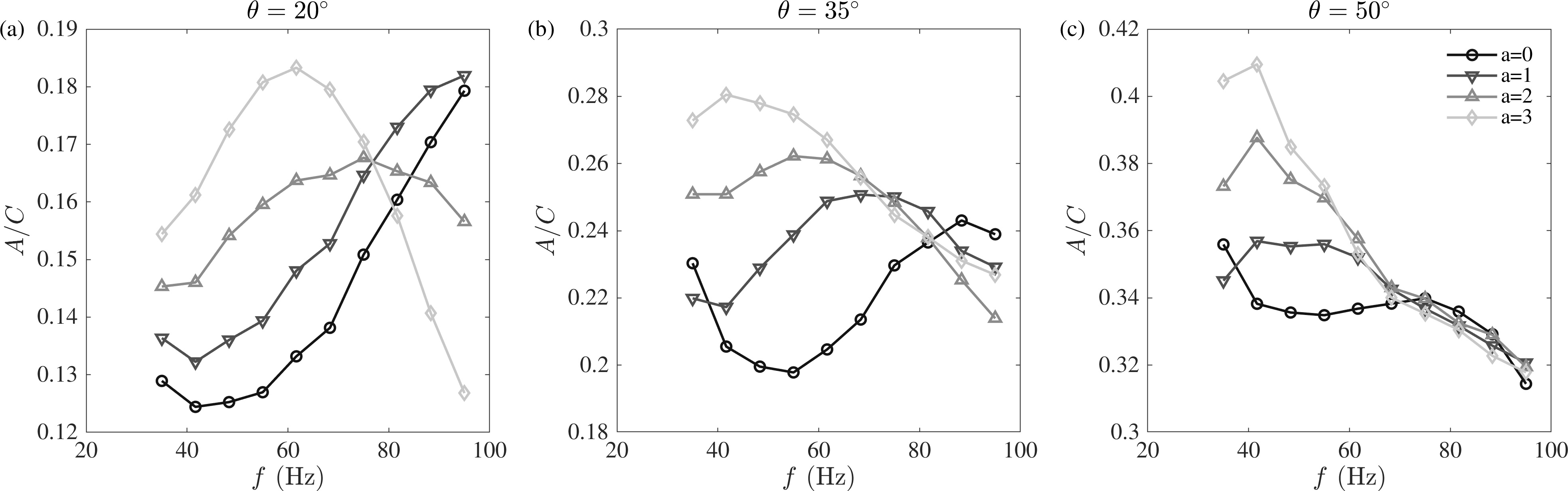}
\caption{Trailing edge amplitude $A$ as a function of frequency for three pitch angles (a) $\theta=20^\circ$, (b) $\theta=35^\circ$ and (c) $\theta=50^\circ$.}
\label{fig:A}
\end{figure}

\begin{figure}[ht!]
\includegraphics[scale=0.15]{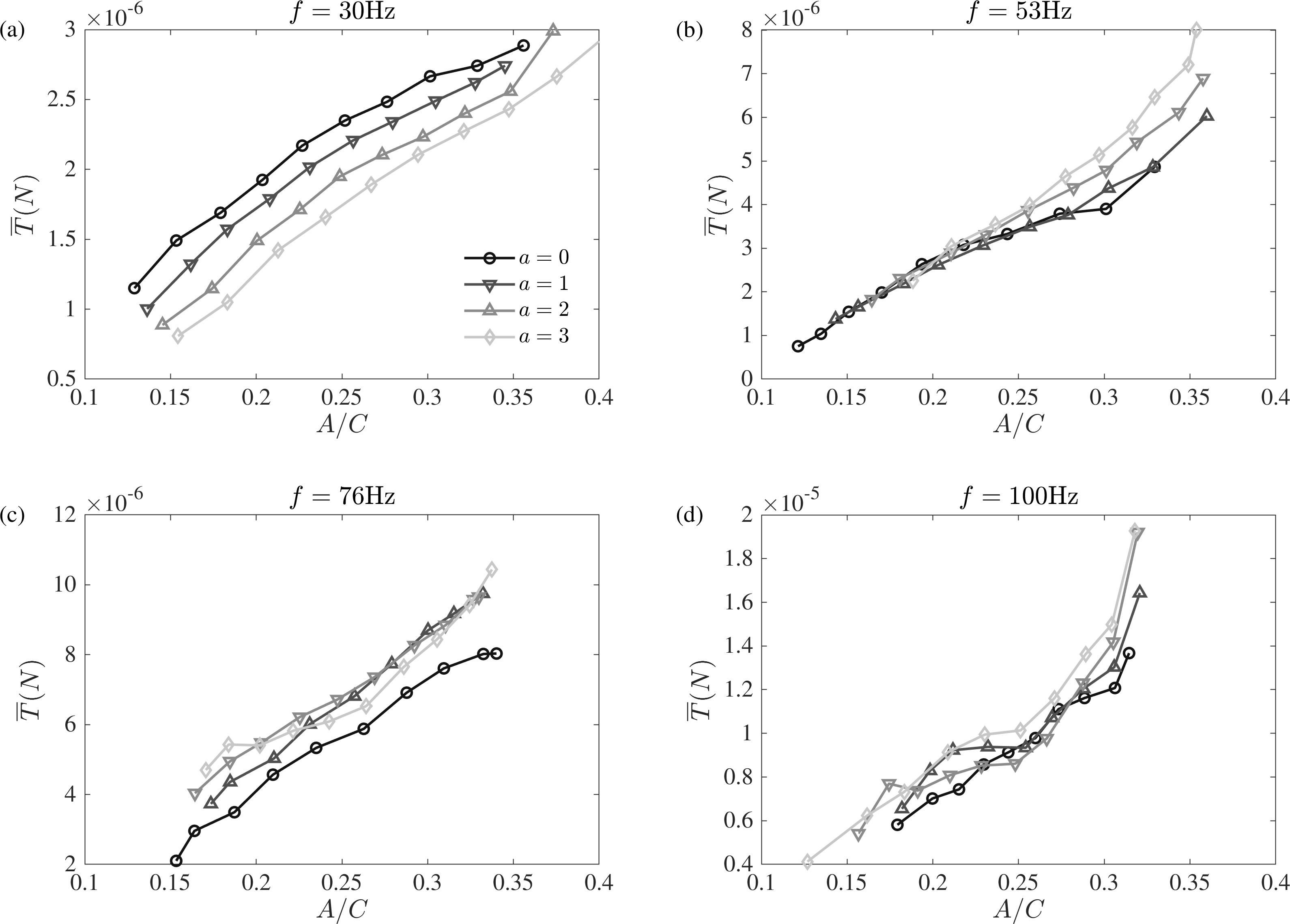}
\caption{Thrust $\overline T$ as a function of tail amplitude $A$ with different pitching frequencies. (a) $f=30$Hz, (b) $f=53$Hz, (c) $f=76$Hz and (d) $f=100$Hz.}
\label{fig:AvsThrust}
\end{figure}

 \subsection{Relation between locomotion and propulsion}

The thrust of a swimmer is obviously related to the swimming motion, which is influenced by the body stiffness.  We look into this relation starting from the amplitude of the tail normalized by the chord length $A/C$. In the current study, the tail motion is the response of the structure to the driven torque on the head, hence the trailing edge amplitude $A$ is an output parameter.  Figure \ref{fig:A} shows the trailing edge amplitude as a function of frequency at different pitching angles. In general, the swimmers with softer tails have larger trailing edge amplitude $A/C$ except for some high frequency cases. However, a soft tail is not advantageous at high frequency domain. For example, the tail amplitude of the soft tail swimmer ($a=3$) drops drastically when $f$ increases beyond $70\mathrm{Hz}$ with a pitching angle of $\theta=20^\circ$. This explains the drop of thrust and swimming speed in Figure \ref{fig:thrust} and Figure \ref{fig:u_vs_a}, respectively.

Previous studies have established a direct correspondence between thrust and trailing edge amplitude \cite[]{saadat2017rules,floryan2018efficient}, since a larger sweeping area of the tail generally indicates more fluid being transported, hence larger thrust. However, as $A$ becomes larger, drag could also increase due to the increase of the angle of attack. This is why the cases with the largest $A/C$ in figure \ref{fig:cot_vs_a} are not exactly the most efficient cases in figure \ref{fig:eta}. Moreover, it is yet to be determined what effect flexibility would have on thrust when the amplitude $A$ is constant.

To examine the relationship between thrust and trailing edge amplitude, we plot thrust $T$ against $A/C$ in figure \ref{fig:AvsThrust}. For all actuation frequencies, the thrust has a quasi-linear relation with the amplitude $A$. Specifically, for higher actuation frequency, the role of posterior flexibility appears to be marginal. At low actuation frequency $f=30$Hz, the uniform stiffness configuration generates more thrust than soft-tail swimmers when the trailing edge amplitude $A$ is the same. 


 \begin{figure}[ht!]
\includegraphics[scale=0.18]{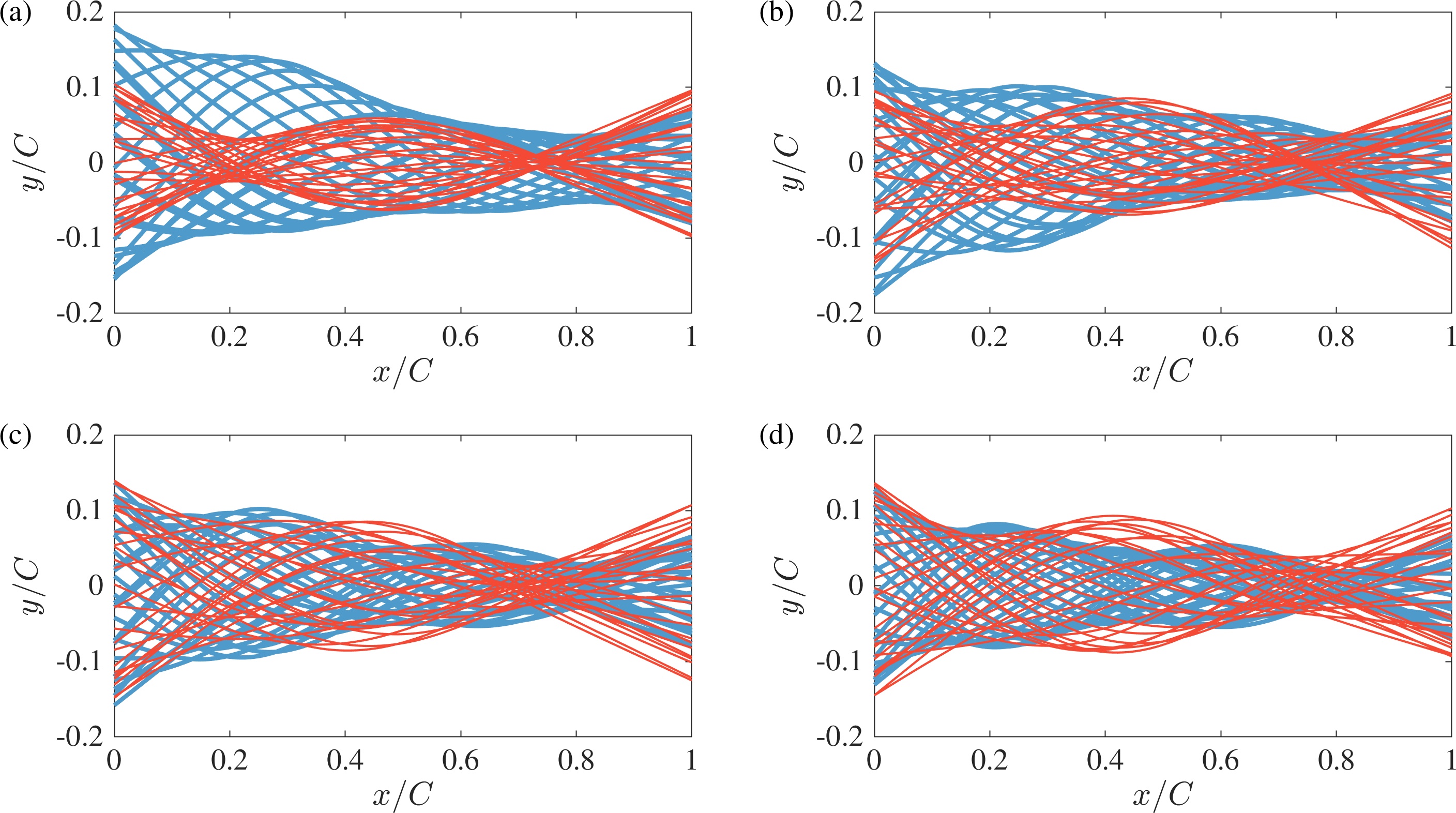}
\caption{The bodylines of fishes aligned at the front end. Panels (a)-(d) correspond to frequencies 60Hz,70Hz,80Hz, and 100Hz, respectively. The red and blue lines represent swimmers with distribution coefficients $a=0$ and $a=3$, respectively.}
\label{fig:midline}
\end{figure}

To compare the locomotion of stiff and soft tail swimmer more clearly, we show the aligned bodyline of the swimmer with a pitching angle of $\theta=20^\circ$ at frequencies 60-100Hz (figure \ref{fig:midline}). Two configurations with stiff tail ($a=0$) and soft tail ($a=3$) are compared. At 60Hz, the soft tail ($a=3$) swimmer has a much larger amplitude in the posterior part than the stiff tail swimmer. In fact, the shape of the bodyline ensemble of the stiff tail swimmer is almost symmetrical at about $x/C=0.5$. This is an indication that the thrust created by the tail is comparable to the resistance of the head, thus resulting in an extremely small total thrust.
As the driving frequency increases, the head-tail symmetry of the stiff tail swimmer is broken. The amplitude of the tail gradually becomes larger than the head. In contrast, the amplitude of the soft tail swimmer gradually diminishes as the frequency increases. Previous studies \cite[]{nguyen2016thrust,quinn2015maximizing} suggest that local maxima in trailing edge amplitude occur when the system is actuated at its natural frequencies. This indicates the frequency is moving away from the natural frequency of the soft tail swimmer, but moving closer to the stiff one. For 100Hz cases, the trailing edge amplitude for the stiff tail swimmer is larger than the soft tail one, giving it more thrust (figure \ref{fig:thrust}a) and speed (figure \ref{fig:u_vs_a}a). 

\begin{figure}[ht!]
\includegraphics[scale=0.18]{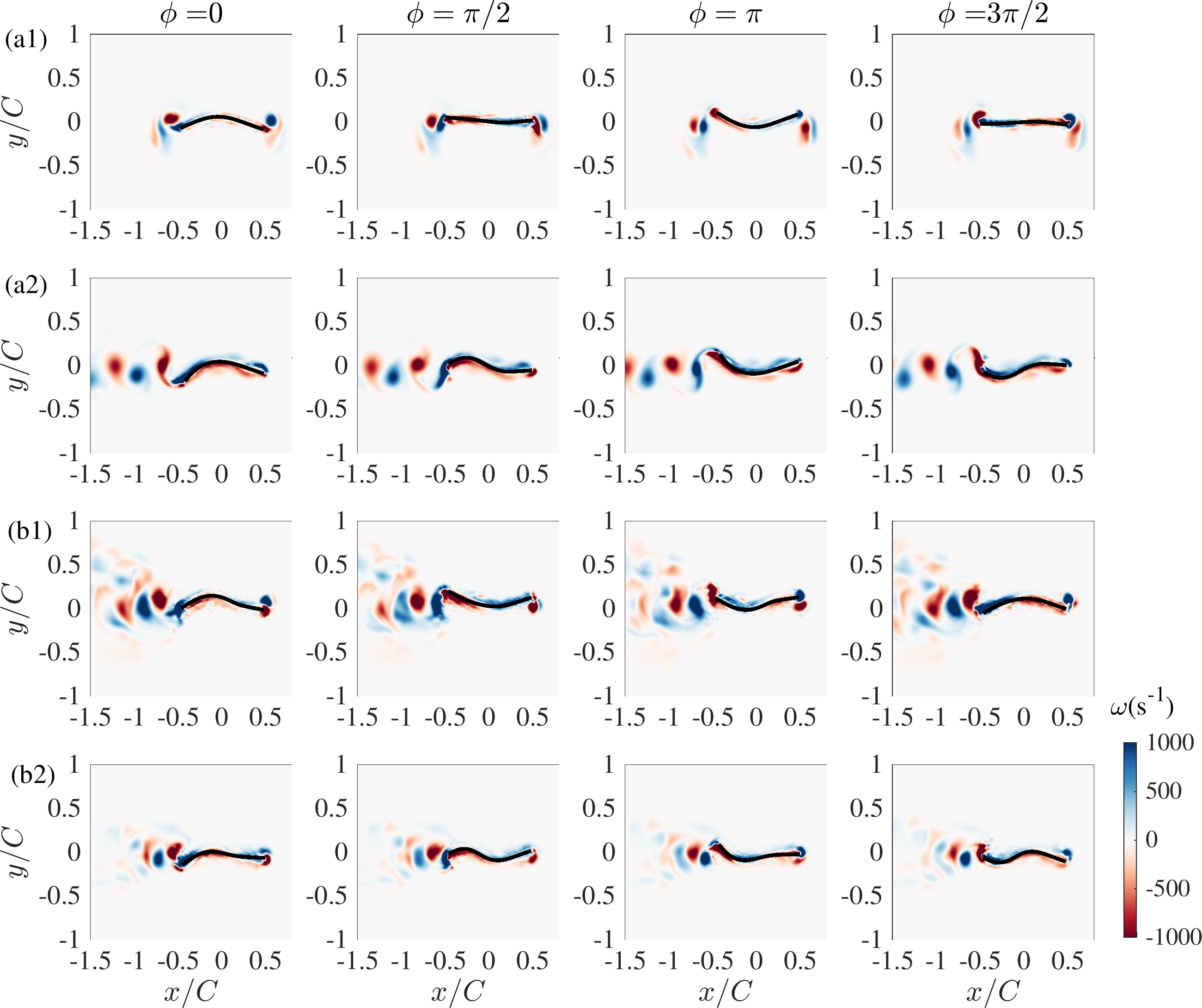}
\caption{Phase-averaged vorticity fields around the swimmers. The pitching angle of all cases is $\theta=20^\circ$. (a1) stiff tail swimmer ($a=0$) at frequency $f=70\mathrm{Hz}$; (a2) soft tail swimmer ($a=3$) at frequency $f=70\mathrm{Hz}$; (b1) stiff tail swimmer ($a=0$) at frequency $f=100\mathrm{Hz}$; (b2) soft tail swimmer ($a=0$) at frequency $f=100\mathrm{Hz}$. To view the vorticity fields for the entire period, please refer to movies 1-4 in the Multimedia material (Multimedia view). }
\label{fig:omega_avg}
\end{figure}

The comparison of the phase-averaged bodyline shapes and surrounding vorticity fields between soft and stiff tail swimmers at driving frequencies of 70Hz and 100Hz is presented in figure \ref{fig:omega_avg}. The corresponding phase-averaged pressure forces of the cases in figure \ref{fig:omega_avg} can be seen in Figure \ref{fig:force_avg}. To observe the evolution of vorticity fields and pressure force of the entire period, refer to movies 1-4 in the Multimedia material (Multimedia view).  At $f=70\mathrm{Hz}$, the amplitude of the front and trailing edge of the stiff tail swimmer ($a=0$) is almost the same, which has already been illustrated in figure \ref{fig:midline}(a). As a result, the thrust generated by the tail is almost completely canceled by the drag on the head (see figure \ref{fig:force_avg}a1, movie 1 (Multimedia view)). The low net thrust leads to a very low swimming speed ($\sim$0.04m/s), and vorticity is mainly located near the front and trailing edge with no vorticity street formed behind the swimmer.  
In contrast, the soft tail swimmer has a much larger trailing edge amplitude and a relatively small amplitude at the front edge at $f=70\mathrm{Hz}$ (see figure \ref{fig:force_avg}a2, movie 2 (Multimedia view)). Consequently, the thrust of the tail is significantly larger while the drag on the head is quite small. In fact, the soft tail swimmer achieves maximal swimming speed ($\sim$ 0.12m/s) and lowest CoT at $f=70\mathrm{Hz}$ when the pitching angle is set $\theta=20^\circ$. Positive and negative vorticity detached from the trailing edge alternatively and formed a trace of vorticity behind the swimmer.

As the actuating frequency increases to $f=100\mathrm{Hz}$, the head-tail symmetry of the uniform stiffness swimmer is broken. A moving wave appears on the posterior part of the body, transporting the fluids towards the wake (figure \ref{fig:force_avg}b1, movie 3 (Multimedia view)). This brings net thrust to the swimmer, which enables it to reach a much higher speed. As a comparison (figure \ref{fig:force_avg}b2, movie 4 (Multimedia view)), the soft tail swimmer appears to have extra fractions of wavelength at the posterior part, indicating a waste of energy on the deformation. In the meantime, the amplitude of the trailing edge also decreases. These changes lead to a decrease in propulsion efficiency and net thrust.

 \begin{figure}[ht!]
\includegraphics[scale=0.15]{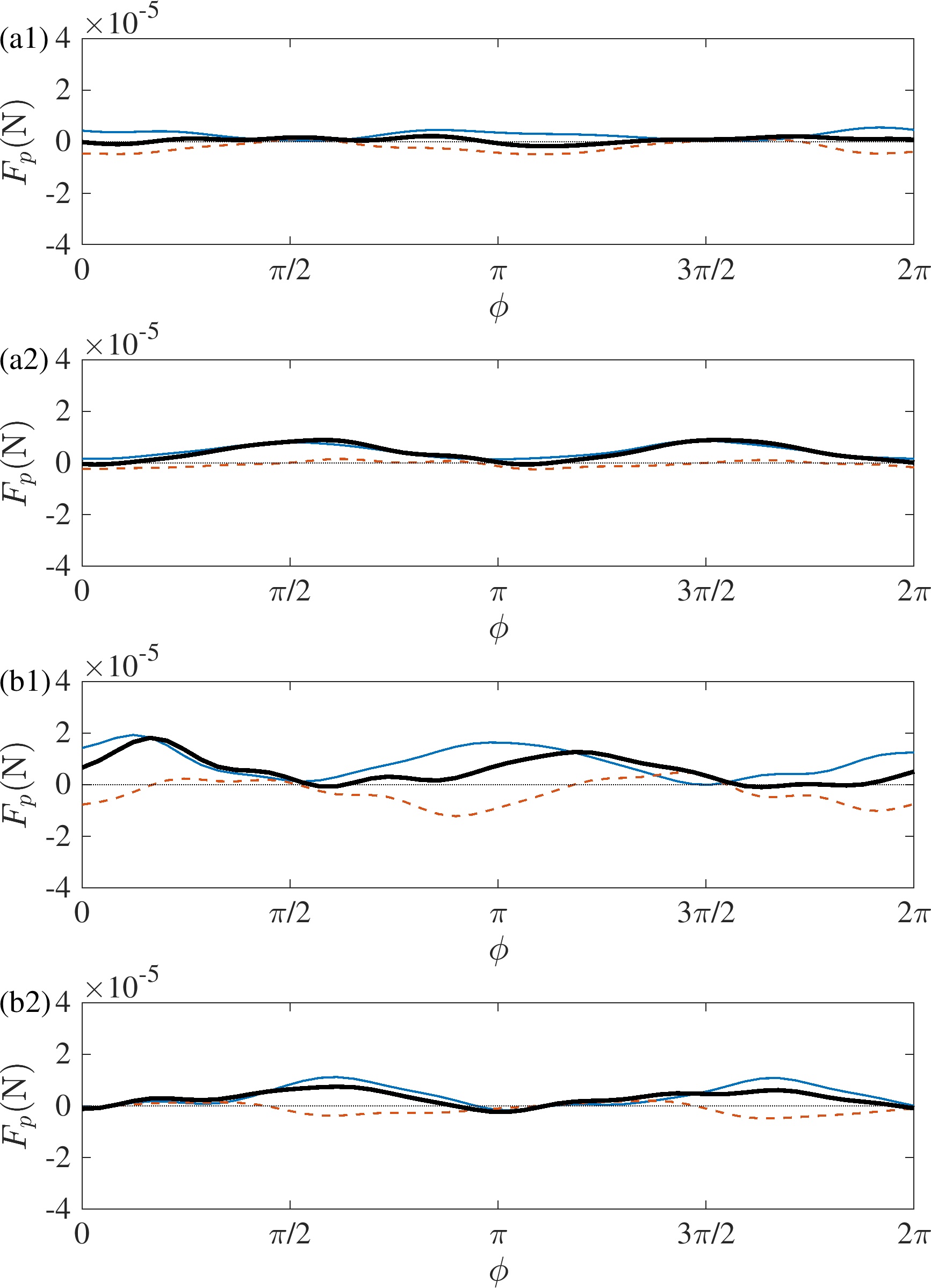}
\caption{Phase-averaged pressure force $F_p$. Blue solid lines indicate the pressure force from the head, red dashed lines indicate the pressure force from the tail and black lines indicate the combined pressure force from the entire body, (a1) $a=0$, $f=70\mathrm{Hz}$, (a2) $a=3$, $f=70\mathrm{Hz}$, (b1) $a=0$, $f=100\mathrm{Hz}$, (b2) $a=3$, $f=100\mathrm{Hz}$. The pitching angle of all cases is $\theta=20^\circ$.}
\label{fig:force_avg}
\end{figure}

\section{Conclusive remarks}\label{sec:conclusion}

 In the present work, we investigate the locomotion performance of a torque-driven swimmer with various stiffness distributions. The complex fluid-structure interaction is simulated with the immersed-boundary method of second-order accuracy within the framework of the Navier-Stokes equations. The space of the input parameters, that is, pitch angle $\theta$, frequency $f$, and distribution coefficient $a$ is explored using the Sobol sequence with more than 600 simulation cases being performed. We confine our input parameters within the bounds of our natural reference, the zebrafish larvae, ensuring consistency. Even with this limited input range, the resulting parameters, such as swimming speed and trailing edge amplitude, cover a broad spectrum, hence a wide variation in Reynolds number $Re$ and Strouhal number $St$ values.

In contrast to most prior research, where a swimmer's movements are preset, our study involves a swimmer that is actuated by a periodic torque applied to its head section. This torque directs the swimmer's head toward specific pitching angles. Apart from this torque, the swimmer's movements, including heaving motions of the leading edge, body undulations, and trailing edge oscillations, result solely from the fluid-structure interaction. This approach allows us to concentrate solely on scenarios of free-swimming that are attainable through real-life actuation methods, e.g. magnetic actuation. Consequently, our current setup bears a closer resemblance to real-world applications of soft-robotic swimmers.

Swimming performance is significantly influenced by the stiffness distribution, particularly when the pitch angle is low or moderate ($\theta\le35^\circ$). In these instances, a sharper decrease trend in the stiffness distribution offers advantages in both swimming speed and energy efficiency compared to a uniform distribution. However, at higher pitch angles, the impact of the stiffness distribution becomes less significant. Further examination of the pressure force reveals that the tail is the source of thrust, while the head accounts for the drag. Furthermore, the correlation between thrust and trailing edge amplitude is confirmed.

It's observed that at a low pitching angle (e.g., $\theta=20^\circ$), for the soft-tail swimmer with $a=3$, an optimal actuation frequency around $f=70$Hz exists. Beyond this point, both swimming speed and propulsion efficiency decline. Detailed examination of the swimmer's movements reveals reduced trailing edge amplitude at higher frequencies and additional deformation. In contrast, the trailing edge amplitude of the swimmer with a uniform stiffness distribution steadily increases with the driving frequency. This could be explained by the differences in their natural frequencies. The soft-tailed swimmer possesses a lower natural frequency compared to the uniform tail swimmer. As the pitch frequency surpasses $70\mathrm{Hz}$ and approaches the natural frequency of the uniform tail swimmer, the soft tail swimmer no longer benefits from resonance, leading to energy waste caused by increased deformation and reduced thrust due to the smaller amplitude of the trailing edge.

The present study offers a clear guideline for applying stiffness distributions in millirobot swimmers. Nevertheless, it is evident that the propulsion efficiency achieved in this study remains considerably lower than the estimated values for natural swimmers\cite[]{van2015body,mueller2004}. In the future, we will explore alternative actuation strategies and stiffness distributions to look for areas to improve.

\section{Acknowledgments}

The study is funded by Deutsche Forschungsgemeinschaft (DFG, German Research Foundation) under Germany’s Excellence
Strategy-EXC2075-390740016. We also thank the Deutsche Forschungsgemeinschaft (DFG, German Research Foundation) for
supporting this work by funding DFG-SFB 1313, Project No.327154368.

\section{Multimedia}

 \begin{figure}[ht!]
\includegraphics[scale=0.5]{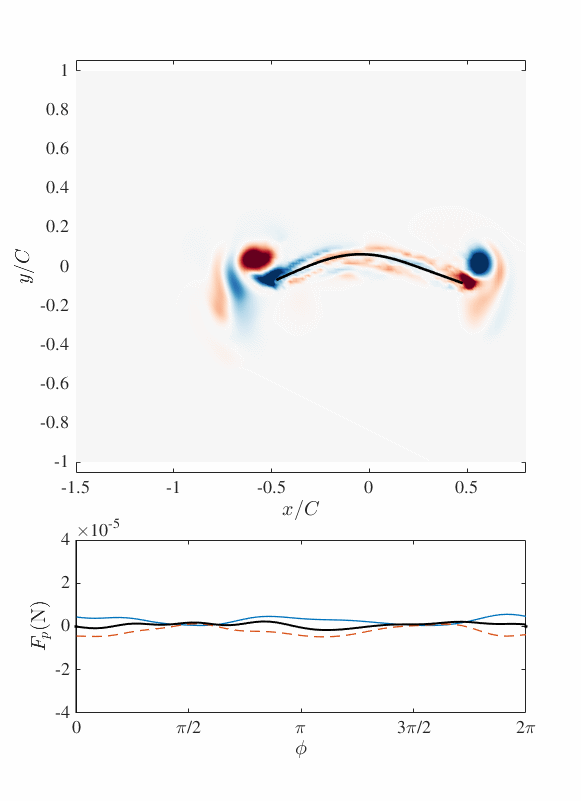}
\caption{Integral Multimedia: movie1}
\label{fig:force_avg}
\end{figure}

 \begin{figure}[ht!]
\includegraphics[scale=0.5]{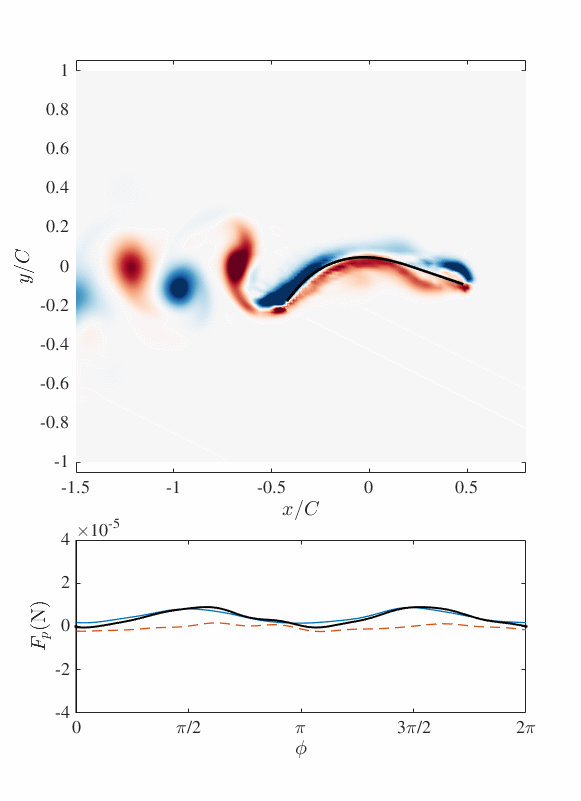}
\caption{Integral Multimedia: movie2}
\label{fig:force_avg}
\end{figure}

 \begin{figure}[ht!]
\includegraphics[scale=0.5]{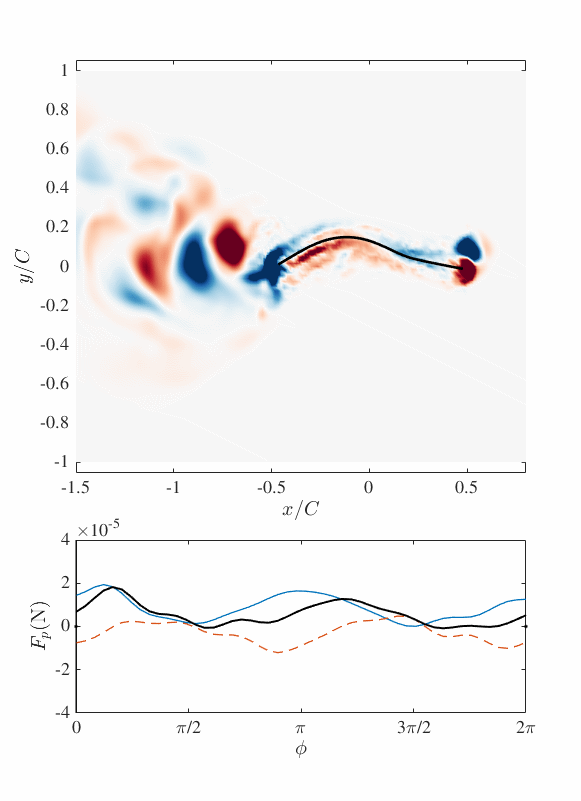}
\caption{Integral Multimedia: movie3}
\label{fig:force_avg}
\end{figure}

 \begin{figure}[ht!]
\includegraphics[scale=0.5]{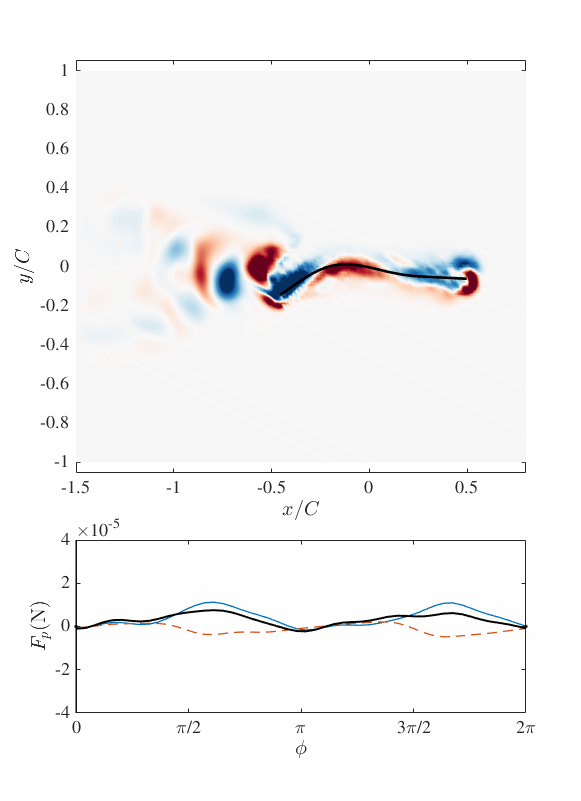}
\caption{Integral Multimedia: movie4}
\label{fig:force_avg}
\end{figure}

\bibliography{references}

\end{document}